\documentclass[aps,floatfix,showpacs,twocolumn]{revtex4} 

\usepackage{graphicx} 
\usepackage{dcolumn} 
\usepackage{amsmath}

\usepackage{graphicx}
\usepackage{longtable}

\begin{document}

\title{Dynamical chiral symmetry breaking and a critical mass}
\author{Lei Chang}
\affiliation{Department of Physics, Peking University, Beijing
100871, China}

\author{Yu-Xin Liu}
\affiliation{Department of Physics, Peking University, Beijing 100871, China} \affiliation{The Key Laboratory of Heavy Ion Physics, Ministry of Education,Beijing
100871, China } 
\affiliation{Center of Theoretical Nuclear Physics, National Laboratory of Heavy Ion Accelerator, Lanzhou 730000, China}

\author{Mandar S.\ Bhagwat}
\affiliation{Physics Division, Argonne National Laboratory, 
             Argonne, IL 60439-4843, U.S.A.} 

\author{Craig D.\ Roberts}
\affiliation{Physics Division, Argonne National Laboratory, 
             Argonne, IL 60439-4843, U.S.A.} 

\author{Stewart V.\ Wright}
\affiliation{Physics Division, Argonne National Laboratory, 
             Argonne, IL 60439-4843, U.S.A.} 

\date{\today}

\begin{abstract}
On a bounded, measurable domain of non-negative current-quark mass, realistic models of QCD's gap equation can simultaneously admit two inequivalent dynamical chiral symmetry breaking (DCSB) solutions and a solution that is unambiguously connected with the realisation of chiral symmetry in the Wigner mode.  The Wigner solution and one of the DCSB solutions are destabilised by a current-quark mass and both disappear when that mass exceeds a critical value.  This critical value also bounds the domain on which the surviving DCSB solution possesses a chiral expansion.  This value can therefore be viewed as an upper bound on the domain within which a perturbative expansion in the current-quark mass around the chiral limit is uniformly valid for physical quantities.  For a pseudoscalar meson constituted of equal mass current-quarks, it corresponds to a mass $m_{0^-} \sim 0.45\,$GeV.  In our discussion we employ properties of the two DCSB solutions of the gap equation that enable a valid definition of $\langle \bar q q \rangle$ in the presence of a nonzero current-mass.  The behaviour of this condensate indicates that the essentially dynamical component of chiral symmetry breaking decreases with increasing current-quark mass.
\end{abstract}


\maketitle

\section{Introduction}
Dynamical chiral symmetry breaking (DCSB) is the creation, via interactions with the gauge field alone, of a fermion mass gap: whose magnitude exceeds, perhaps by a great amount, the mass-scale in the action set by the fermion's bare mass; and which persists when that bare-mass-scale vanishes, namely, in the chiral limit.  It is fundamentally important in strong interaction physics.  For example, DCSB is responsible for the generation of large constituent-like masses for dressed-quarks in QCD, an outcome that could have been anticipated from Refs.\,\cite{lane,politzer}; it is a longstanding prediction of Dyson-Schwinger equation (DSE) studies \cite{cdragw} and has recently been observed in numerical simulations of lattice-regularised QCD \cite{bowman2,bowman}.  DCSB is also the keystone in the realisation of Goldstone's theorem through pseudoscalar mesons in QCD \cite{mrt98}, and thereby the remarkably small value of the ratio of $\pi$- and $\rho$-meson masses, and the weak $\pi\pi$ interaction at low energies \cite{pennington}.  

A large body of efficacious QCD phenomenology is built on an appreciation of the importance of DCSB.  That is evident in studies based on four-fermion interaction models \cite{weisenjl,klevanskynjl,ebertnjl,tandyrev,cahillrev} and in DSE applications \cite{marisrev,schmidtrev,alkoferrev}.  Nevertheless, not all facets of DCSB have been elucidated.  Herein we describe novel aspects of the interplay between explicit and dynamical chiral symmetry breaking.

\section{Dynamical chiral symmetry breaking}
\label{secdcsb}
DCSB can be explored via the gap equation; viz., the DSE for the dressed-fermion self-energy, which for a given quark flavour in QCD is expressed \cite{fnEuclidean}
\begin{eqnarray}
S(p)^{-1} & =&  Z_2 \,(i\gamma\cdot p + m^{\rm bm}) + \Sigma(p)\,, \label{gendse} \\
\Sigma(p) & = & Z_1 \int^\Lambda_q\! g^2 D_{\mu\nu}(p-q) \frac{\lambda^a}{2}\gamma_\mu S(q) \Gamma^a_\nu(q,p) , \label{gensigma}
\end{eqnarray}
where $\int^\Lambda_q$ represents a Poincar\'e invariant regularisation of the integral, with $\Lambda$ the regularisation mass-scale \cite{mrt98,mr97}, $D_{\mu\nu}(k)$ is the dressed-gluon propagator, $\Gamma_\nu(q,p)$ is the dressed-quark-gluon vertex, and $m^{\rm bm}$ is the quark's $\Lambda$-dependent bare current-mass.  The quark-gluon-vertex and quark wave function renormalisation constants, $Z_{1,2}(\zeta^2,\Lambda^2)$, depend on the renormalisation point, $\zeta$, the regularisation mass-scale and the gauge parameter.  

The solution of the gap equation can be written in the following equivalent forms:
\begin{eqnarray} 
\nonumber 
 S(p) & =&  \frac{1}{i \gamma\cdot p \, A(p^2,\zeta^2) + B(p^2,\zeta^2)} 
= \frac{Z(p^2,\zeta^2)}{i\gamma\cdot p + M(p^2)} \\
& = &  - i \gamma\cdot p \,\sigma_V(p^2,\zeta^2) + \sigma_S(p^2,\zeta^2) \,.
\label{Sgeneral}
\end{eqnarray} 
(NB.\ The mass function, $M(p^2)=B(p^2,\zeta^2)/A(p^2,\zeta^2)$, is independent of the renormalisation point.)  It is obtained from Eq.\,(\ref{gendse}) augmented by the renormalisation condition
\begin{equation}
\label{renormS} \left.S(p)^{-1}\right|_{p^2=\zeta^2} = i\gamma\cdot p +
m(\zeta)\,,
\end{equation}
where $m(\zeta)$ is the renormalised (running) mass: 
\begin{equation}
Z_2(\zeta^2,\Lambda^2) \, m^{\rm bm}(\Lambda) = Z_4(\zeta^2,\Lambda^2) \, m(\zeta)\,,
\end{equation}
with $Z_4$ the Lagrangian-mass renormalisation constant.  In QCD the chiral limit is strictly and unambiguously defined by \cite{mrt98,mr97}
\begin{equation}
\label{limchiral}
Z_2(\zeta^2,\Lambda^2) \, m^{\rm bm}(\Lambda) \equiv 0 \,, \forall \Lambda \gg \zeta \,,
\end{equation}
which states that the renormalisation-point-invariant current-quark mass $\hat m = 0$.

QCD's action is chirally invariant in the chiral limit.  Consider a global chiral transformation applied to one particular flavour of quark, characterised by an angle $\theta$.  Under this operation the quark's propagator is modified:
\begin{equation}
S(p) \rightarrow {\rm e}^{i \theta \gamma_5} S(p) {\rm e}^{i \theta \gamma_5} 
= -i\gamma \cdot p \, \sigma_V(p^2) + {\rm e}^{i 2 \theta \gamma_5} \sigma_S(p^2).
\label{chiralT}
\end{equation}
Suppose DCSB takes place so that $B(p^2,\zeta^2) \not \equiv 0$.  Then, with the choice $\theta = \pi/2$, Eq.\,(\ref{chiralT}) corresponds to mapping $B(p^2,\zeta^2) \to - B(p^2,\zeta^2)$.  It follows that if $B(p^2,\zeta^2)$ is a solution of the gap equation in the chiral limit, then so is $[- B(p^2,\zeta^2)]$.  While these two solutions are distinct, the chiral symmetry entails that each yields the same pressure \cite{fnpressure}.  Hence they correspond to equivalent vacua.  This is an analogue of the chiral-limit equivalence between the $(\sigma =1,\pi=0)$ and \mbox{$(\sigma =-1,\pi=0)$} vacua in the linear-sigma-model, as elucidated in Refs.\,\cite{reggcm,cdrqed}.  It is notable that, more generally, given a solution of the $m(\zeta) > 0$ gap equation characterised by $\{A_{m(\zeta)}(p^2,\zeta^2),B_{m(\zeta)}(p^2,\zeta^2)\}$, then $\{A_{-m(\zeta)}(p^2,\zeta^2),-B_{-m(\zeta)}(p^2,\zeta^2)\}$ is a solution of the gap equation obtained with $[-m(\zeta)]$.

Studies of DCSB have hitherto focused on a positive definite solution of the gap equation because the introduction of a positive current-quark bare-mass favours this solution; viz., if another solution exists, then it has a lower pressure.  Returning again to the sigma-model analogy, such a bare-mass tilts the so-called wine-bottle potential, producing a global minimum at $(\sigma =1,\pi=0)$.  However, whether the massive gap equation admits solutions other than that which is positive definite, the effect of the current-quark mass on such solutions, and their interpretation, are questions little considered.

\section{Exemplar}
\label{secnjl}
To begin addressing these questions, we first consider the simple example defined by Eqs.\,(\ref{gendse}), (\ref{gensigma}) with the following forms for the dressed-gluon propagator and quark-gluon vertex:
\begin{eqnarray}
\label{Dnjl}
g^2 D_{\mu\nu}(p-q) & = &  \delta_{\mu\nu}\, 
\frac{1}{m_G^2}\,\theta(\tilde\Lambda^2-q^2)\,,\\
\label{Gnjl}
\Gamma_\nu^a(q,p) & = & \gamma_\nu \frac{\lambda^a}{2}\,,
\end{eqnarray}
wherein $m_G$ is some ``gluon'' mass-scale and $\tilde\Lambda$ serves as a cutoff \cite{fn1}.  The model thus obtained is not renormalisable so that the regularisation scale $\tilde\Lambda$, upon which all calculated quantities depend, plays a dynamical role and the renormalisation constants can be set to one.  In the model thus defined the gap equation is
\begin{eqnarray} 
\nonumber \lefteqn{ 
i\gamma\cdot p\, A(p^2) + B(p^2)= i \gamma\cdot p + m^{\rm bm} }\\ 
\nonumber &+ & 
 \frac{4}{3}\, \frac{1}{m_G^2}\,\int \frac{d^4q}{(2\pi)^4}\, 
\theta(\tilde\Lambda^2 -q^2)\,\\
&& \times  \gamma_\mu \,\frac{-i\gamma\cdot q A(q^2) + 
B(q^2)}{q^2 A^2(q^2) + B^2(q^2)}\, \gamma_\mu\,.
 \label{njlgap2} 
\end{eqnarray} 

This gap equation's solution is $A(p^2) \equiv 1$ and \mbox{$B(p^2) = M$}, a constant which satisfies 
\begin{eqnarray} 
\label{njlGE}
M & = &  m^{\rm bm} + M\,\frac{1}{3\pi^2} \, \frac{1}{m_G^2}\, { C}(M^2,\tilde\Lambda^2)\,,\\ 
\label{CMLamda} { C}(M^2,\tilde\Lambda^2) & = & \tilde\Lambda^2 - M^2 \ln\left[1+\tilde\Lambda^2/M^2\right]
\,. 
\end{eqnarray} 
Since $\tilde\Lambda$ defines the mass-scale in a nonrenormalisable model, we can set $\tilde\Lambda \equiv 1$ and hereafter interpret all other mass-scales as being expressed in units of $\tilde\Lambda$, whereupon the gap equation becomes 
\begin{equation} 
\label{pgapnjl} G(M):= M - m^{\rm bm} - M\,\frac{1}{3\pi^2} \, \frac{1}{m_G^2}\, { 
C}(M^2,1) = 0\,. 
\end{equation} 

Equation~(\ref{pgapnjl}) admits a $M\neq 0$ solution when $m^{\rm bm}=0$ if and only if \begin{equation}
\label{critmG1}
m_G^2 < (m_G^{\rm cr})^2 = \frac{1}{3\pi^2}\,;
\end{equation}
namely, it supports DCSB in this case.  Hence, to proceed we choose
\begin{equation}
\label{usedmG1}
m_G^2 = \frac{3}{4}\, \frac{1}{3\pi^2}\,.
\end{equation}
NB.\ For $m_G > m_G^{\rm cr}$ the only solution of the gap equation is one that may be obtained via a perturbative expansion in the coupling and hence DCSB is impossible.  That domain is therefore not of interest herein.

\begin{figure}[t]
\centerline{
\includegraphics[width=0.45\textwidth]{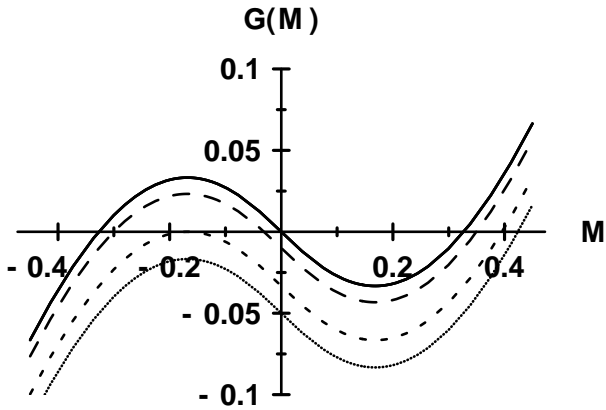}}
\vspace*{2ex}

\caption{\label{figNJL} The zeros of $G(M)$ give the solution of the gap equation defined by Eqs.\,(\protect\ref{pgapnjl}), (\protect\ref{usedmG1}).
\textit{Solid curve}: obtained with $m^{\rm bm}=0$, in which case $G(M)$ is odd under $M\to -M$; \textit{long-dashed curve}: $m^{\rm bm}=0.01$; \textit{short-dashed curve}: $m^{\rm bm}=m_{\rm cr}^{\rm bm} = 0.033$; \textit{dotted curve}: $m^{\rm bm}=0.05$.
(All dimensioned quantities in units of $\tilde\Lambda$.)}
\end{figure}

$G(M)$ is plotted in Fig.\,\ref{figNJL}.  One reads from the figure that in the chiral limit there are three solutions to the gap equation: 
\begin{equation}
\label{gapsolnjl}
M = \left\{
\begin{array}{l}
M_W = 0\,, \\
M_\pm = \pm M^0 = \pm 0.33\,.
\end{array}\right.
\end{equation}
$M_W$ describes a realisation of chiral symmetry in the Wigner mode.  It corresponds to the vacuum configuration in which the possibility of DCSB is not realised; i.e., $\sigma = 0 = \pi$ in the sigma-model analogy.  In the chiral limit this is the only solution accessible via a perturbative expansion in the coupling.  The solutions $M_\pm$ are essentially nonperturbative.  They represent the realisation of chiral symmetry in the Nambu-Goldstone mode; namely, DCSB.

It is apparent in Fig.\,\ref{figNJL} that with the gap equation obtained via Eqs.\,(\ref{Dnjl}), (\ref{Gnjl}), each of the solutions identified in Eq.\,(\ref{gapsolnjl}) evolves smoothly with current-quark mass on a neighbourhood of $m^{\rm bm}=0$.  
We subsequently consider the manner in which these solutions evolve as $m^{\rm bm}$ is increased from zero.  In that discussion we will retain the labels introduced in Eq.\,(\ref{gapsolnjl}) and attach them to that solution which is (pointwise, if relevant) closest in magnitude to the chiral limit solution of the same name.

$M=M_+>0$ is the solution usually tracked in connection with QCD phenomenology.  In models of this type it is identified as a constituent-quark mass.  As $m^{\rm bm}$ is increased, $M_+$ also increases.  

As evident in Fig.\,\ref{figNJL}, the other two solutions of the gap equation do not immediately disappear when $m^{\rm bm}$ increases from zero.  Nor do they always persist.  Instead, these solutions exist on a domain 
\begin{equation}
{\cal D}(m^{\rm bm})= \{m^{\rm bm}\; | \;0 \leq m^{\rm bm} < m^{\rm bm}_{\rm cr} \}.
\end{equation}
$M_W,M_-$ also evolve smoothly with $m^{\rm bm}$.  Moreover, at the critical current-quark mass, $m^{\rm bm}_{\rm cr}$, these two solutions coalesce.  

To understand the origin of a critical mass we observe from Eq.\,(\protect\ref{pgapnjl}) and Fig.\,\ref{figNJL} that introducing a current-quark mass merely produces a constant pointwise negative shift in the curve $G(M)$.  Hence, the critical current-quark mass is that value of this mass for which $G(M)=0$ at its local maximum.  The local maximum occurs at 
\begin{equation}
M_{\rm lm} = \{M \;|\; M<0\,,\; G^\prime(M)\} = 0
\end{equation}
and therefore in the present illustration the critical current-quark mass 
\begin{equation}
\label{mbmcr}
m^{\rm bm}_{\rm cr} = \{ m^{\rm bm} \; | \;G(M_{\rm lm}) = 0 \} = 0.033 \,.
\end{equation}

\begin{figure}[t]
\centerline{
\includegraphics[width=0.45\textwidth]{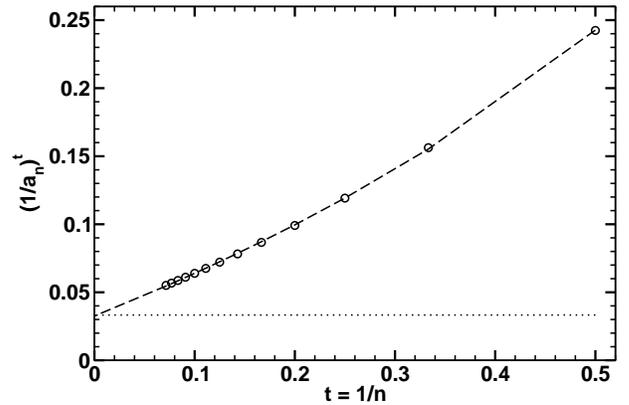}}
\vspace*{2ex}

\caption{\label{Mpm} \textit{Circles} -- Reciprocal of the $n$-th roots of the coefficients in a chiral expansion of $M_+$ around $m^{\rm bm}=0$, Eq.\,(\protect\ref{Mpchiral}), as a function of $t=1/n$. \textit{Dashed curve} -- the function in Eq.\,(\protect\ref{fitMp}).  \textit{Dotted curve} -- $m^{\rm bm}_{\rm cr} = 0.033$. 
(All dimensioned quantities in units of $\tilde\Lambda$.)}
\end{figure}

We now return to the behaviour of $M_+$ and ask whether this quantity has a power series expansion in $m^{\rm bm}$ about $m^{\rm bm}=0$; viz., a \emph{chiral} expansion:
\begin{equation}
\label{Mpchiral}
M_+(m^{\rm bm}) = M_0 + \sum_{n=1}^{\infty} a_n (m^{\rm bm})^n\,.
\end{equation}
Such an expansion exists; i.e., is absolutely convergent on a measurable domain, so long as $\,\forall\,n$
\begin{equation}
\label{sequence}
\left(\frac{1}{|a_n|}\right)^{1/n} > \left(\frac{1}{|a_{n+1}|}\right)^{1/(n+1)} 
\end{equation}
and
\begin{equation}
\label{mrc}
m_{\rm rc}:= \lim_{n\to \infty} \left(\frac{1}{|a_n|}\right)^{1/n} > 0 \,,
\end{equation}
where the quantity $m_{\rm rc}$ is the radius of convergence for the series; i.e., the series converges on $0 \leq m^{\rm bm} <m_{\rm rc}$.

To determine whether $M_+$ has such an expansion we inserted Eq.\,(\ref{Mpchiral}) into Eq.\,(\ref{pgapnjl}) and solved the sequence of algebraic equations that this produces to obtain the coefficients $\{a_1,\ldots,a_{14}\}$.  The procedure is straightforward but we stopped at $n=14$ because the magnitude of the coefficients grows rapidly with $n$; e.g., $a_{14}= -4.27331\times 10^{17}$, and this order was sufficient for our purpose.  The information depicted in Fig.~\ref{Mpm} indicates that $M_+$ does have a chiral expansion with a nonzero but finite radius of convergence.  The curve in the figure is the function ($m=1$ diagonal Pad\'e)
\begin{equation}
\label{fitMp}
u_0 + \frac{u_1 t}{1+u_2 t}\,,\; u_0=0.326\,,\;u_1=0.295\,,\;u_2= -0.596\,,
\end{equation}
where the coefficients were fixed in a least-squares fit to $\{(1/a_n)^{1/n},n=2,\ldots,14\}$.  The limit $t\to 0$ corresponds to $n\to \infty$ and hence the value of $u_0$ gives the radius of convergence.  We repeated the fit with diagonal Pad\'e approximants of order $m=2,3,4$.  Combining the results, we obtain the radius of convergence for the chiral expansion of $M_+$:
\begin{equation}
\label{radiusc}
m_{\rm rc} = 0.034\pm 0.001 = m^{\rm bm}_{\rm cr}
\end{equation}
within numerical error.  Hence ${\cal D}(m^{\rm bm})$ is also the domain on which $M_+$ has a chiral expansion.  (NB.\ Analogous chiral expansions exist within this domain for $M_W$, $M_-$.  The expansion for $M_W$ vanishes for $m^{\rm bm}=0$ whereas the leading term in that for $M_-$ is $M_0$.  We emphasise that $M_\pm$ have chiral expansions around the solution that is \emph{nonperturbative} in the coupling.  A nonzero value of $M_0$ is never attainable via an expansion in the coupling.)

\begin{figure}[t]
\centerline{
\includegraphics[width=0.48\textwidth]{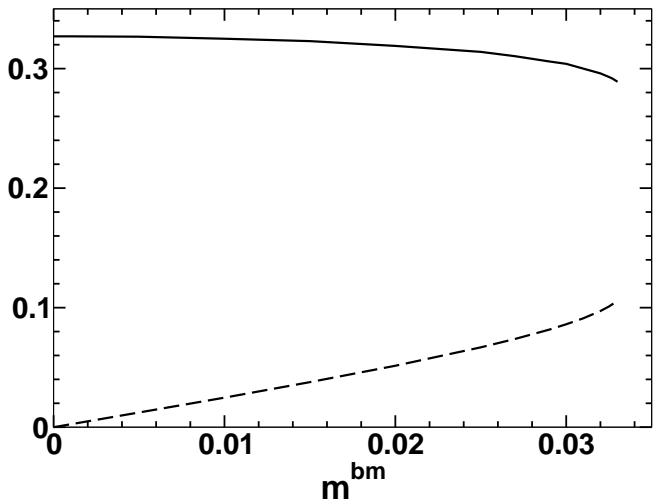}}
\vspace*{2ex}

\caption{\label{figbarM} Evolution with current-quark mass of $\bar M$ (solid curve) and $\check M$ (dashed curve).
(All dimensioned quantities in units of $\tilde\Lambda$.)}
\end{figure}

The two combinations 
\begin{equation}
\label{MbMc}
\bar M := \frac{1}{2} \left(M_+ - M_-\right)\,,\; \check M := \frac{1}{2} \left(M_+ + M_-\right)
\end{equation}
are of interest.  In the chiral limit, $\bar M = M^0$ and $\check M = M_W$.  The evolution of each with current-quark mass is depicted in Fig.\,\ref{figbarM}.  It is apparent that $\bar M$ is continuous on ${\cal D}(m^{\rm bm})$ and evolves from the DCSB solution with increasing $m^{\rm bm}$: $\bar M(m^{\rm bm})$ is a monotonically decreasing function.  These features can be understood as illustrating that the essentially dynamical component of chiral symmetry breaking decreases with increasing current-quark mass.  This has also been argued via the constituent-quark $\sigma$-term; e.g., in Sec.\,5.2.2 of Ref.\,\cite{HUGS05}.

The alternative combination, $\check M$, is also continuous on ${\cal D}(m^{\rm bm})$.  With the value of the coupling given in Eq.\,(\ref{usedmG1}), $\check M(m^{\rm bm})$ evolves from the Wigner solution according to
\begin{equation}
\check M \stackrel{m^{\rm bm} \sim 0}{=}  m^{\rm bm}\left[1 + \frac{2}{3}\, \frac{3 (M^0 )^2 - 1 }{(M^0 )^2+ 1 }\right]+ \ldots\ \;.
\end{equation}
The development of $\check M(m^{\rm bm})$ might be viewed as a gauge of the destabilising effect that DCSB has on this model's Wigner mode.

We now return to the critical current-quark mass, Eq.\,(\ref{mbmcr}), (\ref{radiusc}).  In the neighbourhood of $m^{\rm bm}=0$, 
\begin{equation}
 M_W(m^{\rm bm}) = - 3 \, m^{\rm bm} + \ldots\ ;
\end{equation}
viz., $M_W(m^{\rm bm})$ can be expressed as a power series in $m^{\rm bm}$ around its chiral limit value, where that value is perturbative in the coupling.  However, with increasing current-quark mass $M_W$ decreases steadily toward $M_-$, which is nonzero in the chiral limit and essentially nonperturbative in the coupling, until at $m^{\rm bm}_{\rm cr}$, $M_W =M_-$.  At this point a solution whose small current-quark mass behaviour is essentially perturbative has melded with a solution that is inaccessible in perturbation theory and actually characteristic of DCSB.  A related view sees $M_-$ as a DCSB solution whose modification by a current-quark mass can no longer be evaluated as a power series in $m^{\rm bm}$ when that mass exceeds $m^{\rm bm}_{\rm cr}$.  Finally, a chiral expansion of $M_+$ can only converge for current-quark masses less than $m^{\rm bm}_{\rm cr}$.  

These observations suggest that $m^{\rm bm}_{\rm cr}$ specifies the upper bound on the domain within which, for physically relevant quantities, a perturbative expansion in the current-quark mass around their chiral limit values can be valid; i.e., it is a (possibly weak) upper bound on the radius of convergence.  This view and the numerical result in Eq.\,(\ref{mbmcr}) coincide with Ref.\,\cite{hatsuda}.  We will return to this point.

\section{Closer to QCD}
It is natural to ask whether analogous behaviour exists in QCD.  To explore this we work with a renormalisation-group-improved (RGI) rainbow truncation of the gap equation's kernel.  This is the leading-order in a systematic and symmetry-preserving truncation of the DSEs that is noperturbative in the coupling \cite{munczek,truncscheme,detmoldvertex,bhagwatvertex}.  The truncation has been used widely; e.g., Refs.\,\cite{mr97,jainmunczek,klabucar}, and references thereto, and an efficacious implementation preserves the one-loop ultraviolet behaviour of perturbative QCD.  However, a model assumption is required for the behaviour of the kernel in the infrared; viz., on the interval $Q^2 \lesssim 1\,$GeV$^2$, which corresponds to length-scales $\gtrsim 0.2\,$fm. 

The rainbow truncation is realised in the gap equation via the replacement of Eq.\,(\ref{gensigma}) by \cite{mr97}
\begin{equation} 
%
\Sigma(p)=\int^\Lambda_q\! {\cal G}((p-q)^2) D_{\mu\nu}^{\rm free}(p-q) \frac{\lambda^a}{2}\gamma_\mu S(q) \frac{\lambda^a}{2}\gamma_\nu . \label{rainbowdse} 
\end{equation} 
Herein we employ the model interaction introduced in Ref.\,\cite{maristandy1}:
\begin{equation}
\frac{{\cal G}(t)}{t} = \frac{4\pi^2}{\omega^6} D\, t {\rm
e}^{-t/\omega^2}
 + \, \frac{ 8\pi^2\, \gamma_m } { \ln\left[\tau + \left(1 +
t/\Lambda_{\rm QCD}^2\right)^2\right]} \, {\cal F}(t) \,, \label{gk2}
\end{equation}
with $t=k^2$, ${\cal F}(t)= [1-\exp(-t/[4 m_{\cal F}^2])]/t$, $m_{\cal F}^2=0.5\,$GeV, $\tau={\rm e}^2-1$, $\gamma_m=12/25$ and $\Lambda_{\rm QCD} = \Lambda^{(4)}_{\overline{MS}} = 0.234$.  This form expresses the interaction as a sum of two terms.  The second ensures that perturbative behaviour is correctly realised at short range; namely, as written, for $(k-q)^2 \sim k^2 \sim q^2 \gtrsim 1 - 2\,$GeV$^2$, Eq.\,(\ref{gk2}) guarantees that the quark-antiquark scattering kernel, $K$, is precisely as prescribed by QCD.  On the other hand, the first term in ${\cal G}(t)$ is a model for the long-range behaviour of the interaction.  It is a finite width representation of the form introduced in Ref.\,\cite{mn83}, which has been rendered as an integrable regularisation of $1/t^2=1/k^4$ \cite{mm97}.  This interpretation, when combined with the result that in a heavy-quark--heavy-antiquark BSE the RGI ladder truncation is exact \cite{bhagwatvertex}, is consistent with ${\cal G}(t)$ leading to a Richardson-like potential \cite{richardson} between static sources.

The true parameters in Eq.\,(\ref{gk2}) are $D$ and $\omega$, which together determine the integrated infrared strength of the rainbow kernel.  However, they are not independent \cite{maristandy1}: in fitting to a selection of ground-state observables, a change in one is compensated by altering the other; e.g., on the interval $\omega\in[0.3,0.5]\,$GeV, the fitted observables are approximately constant along the trajectory \cite{raya3}
\begin{equation}
\label{omegaD}
\omega \,D = (0.72\,{\rm GeV})^3.
\end{equation}
Equation (\ref{gk2}) is thus a one-parameter model.  

It is important to bear in mind that because the truncation preserves the one-loop renormalisation group properties of QCD the ultraviolet behaviour of the solutions of Eqs.\,(\ref{gendse}) and (\ref{rainbowdse}) is precisely that of QCD.  Hence we have 
\begin{equation}
\label{Mp2uv}
M(p^2) \stackrel{p^2\gg \Lambda^2_{\rm QCD}}{=} \frac{\hat m}{(\mbox{\small $\frac{1}{2}$} \ln p^2/\Lambda^2_{\rm QCD})^{\gamma_m}},
\end{equation}
where $\hat m$ is the renormalisation-group-invariant mass.  

The model-dependence is mainly restricted to infrared momenta but on this domain, too, there is good agreement with QCD; e.g., the gap equation solutions are in semiquantitative agreement \cite{raya3} with numerical simulations of lattice-regularised quenched-QCD.  (NB.\ Precise agreement would be incorrect because Eq.\,(\ref{gk2}) corresponds to an unquenched theory.) The conditions have been explored under which pointwise agreement between DSE results and quenched- and full-lattice simulations may be obtained \cite{bhagwat,bhagwat2,alkoferdetmold}.

Equations~(\ref{gendse}) and (\ref{rainbowdse}) obviously admit the \mbox{$M(p^2)\equiv 0$} solution in the chiral limit, Eq.\,(\ref{limchiral}).  This solution, which can always be obtained through a weak coupling expansion, is analogous to $M_W$ in Eq.\,(\ref{gapsolnjl}).  

In the chiral limit the rainbow gap equation also yields a DCSB solution.  This capacity is the basis for much of the phenomenological success of the RGI rainbow-ladder truncation.  The truncation preserves the feature that if $M_+(p^2)=M(p^2)>0$, $\forall p^2>0$, is a solution of the chiral-limit gap equation, then so is $M_-(p^2):= [-M(p^2)]$.  These solutions are the analogues of $M_\pm$ in Eq.\,(\ref{gapsolnjl}).  

We solve the gap equation using the Pauli-Villars regularisation procedure described in Ref.\,\cite{empirad} and work in Landau gauge because it is a fixed point of the renormalisation group.  The renormalisation condition, Eq.\,(\ref{renormS}), is implemented at $\zeta = \zeta_{19}:=19\,$GeV with a choice for the current-quark mass.  This scale is deep in the perturbative (weak-coupling) domain and hence a renormalisation-group-invariant current-quark mass is unambiguously specified via Eq.\,(\ref{Mp2uv}).  For reference, we also list below current-quark masses defined therewith via one-loop evolution to a ``typical hadron scale''; viz., 
\begin{equation}
m(\zeta_1) := \frac{\hat m}{(\ln \zeta_1/\Lambda_{\rm QCD})^{\gamma_m}} \,, \; \zeta_1=1\,{\rm GeV}\,.
\end{equation}

\begin{figure}[t]
\centerline{\includegraphics[width=0.50\textwidth]{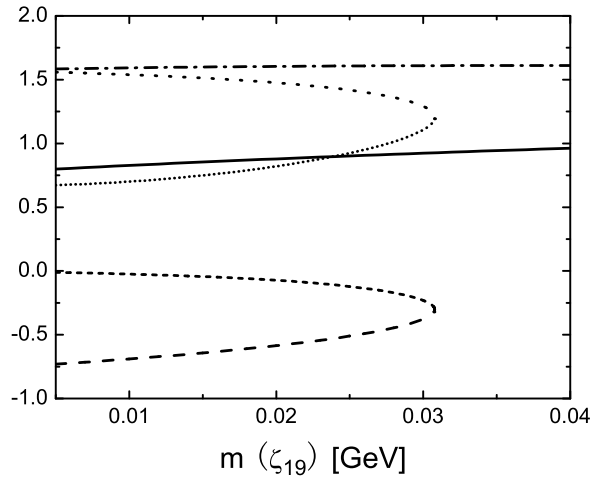}}

\caption{\label{MTAB} Evolution with current-quark mass, $m(\zeta_{19})$, of \mbox{$A(p^2=0,\zeta^2)$} (dimensionless), $B(p^2=0,\zeta^2)$ (GeV) as calculated with $\omega=0.4\,$GeV in Eq.\,(\protect\ref{gk2}): \textit{solid curve} -- $B_+(0)$; \textit{dash-dot curve} -- $A_+(0)$; \textit{long dashed curve} -- $B_-(0)$; \textit{sparse dotted curve} -- $A_-(0)$; \textit{short dashed curve} -- $B_W(0)$; \textit{dense dotted curve} -- $A_W(0)$.  
}
\end{figure}

It is apparent in Fig.\,\ref{MTAB} that this interaction model, too, exhibits a bounded domain of current-quark mass on which $M_-(p^2)$ and $M_+(p^2)$ exist simultaneously:
\begin{equation}
{\cal D}(\hat m)= \{\hat m \; | \;0 \leq \hat m < \hat m_{\rm cr} \}\,.
\end{equation}
Away from the chiral limit the solution characterised by $M_-(p^2)$ describes the propagation characteristics of a quark embedded in an unstable vacuum.  The properties of that vacuum may be detailed, e.g., by solving the Bethe-Salpeter equation for meson bound-states with this dressed-quark propagator.  The critical current-quark mass depends on $\omega$:
\begin{equation}
\label{MTmcr}
\begin{array}{l|ccc}
\omega \, [{\rm GeV}] & 0.3 & 0.4 & 0.5 \\\hline
\rule{0ex}{2.5ex}\hat m_{\rm cr} \, [{\rm MeV}] & 71 & 63 & 31\\\hline
\rule{0ex}{2.5ex} m_{\rm cr}(\zeta_{19}) \, [{\rm MeV}] & 35 & 31 & 15\\
\rule{0ex}{2.5ex} m_{\rm cr}(\zeta_1) \, [{\rm MeV}] & 60 & 53 & 26
\end{array}\,,
\end{equation}
where the third and fourth rows report the critical mass at the renormalisation scales described above.  (NB.\ In most phenomenological applications $0.3 < \omega < 0.4$.)  In analyses of hadron observables founded on the models that form the basis of our arguments, the $s$-quark RGI current-mass exceeds $\hat m_{\rm cr}$.


\begin{figure}[t]
\centerline{\includegraphics[width=0.5\textwidth]{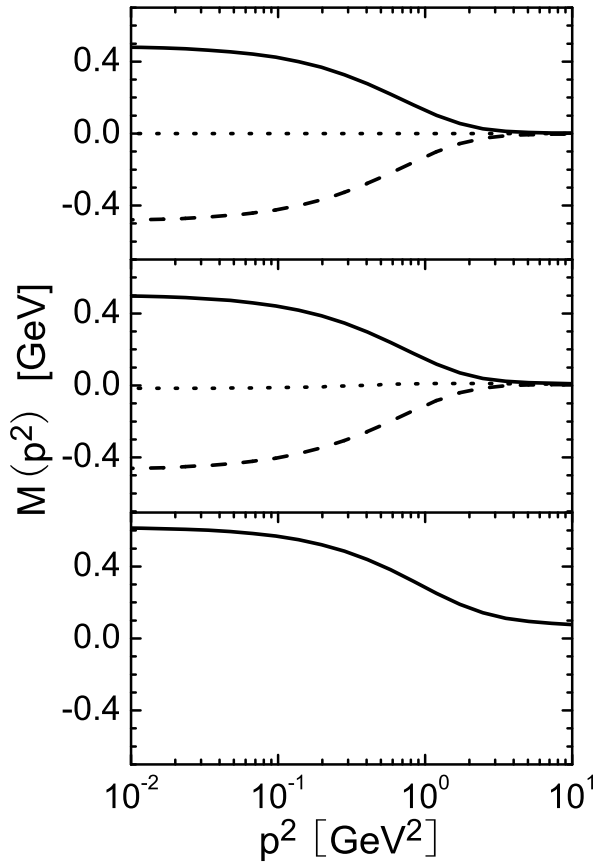}}

\caption{\label{figMpmp2} Momentum dependence of the dressed-quark mass-function, $M(p^2)$: \textit{upper panel} -- chiral limit; \textit{middle panel} -- $m(\zeta_{19})=5\,$MeV; \textit{lower panel} -- $m(\zeta_{19})=50\,$MeV, for which there is naturally no $M_-(p^2)$ solution.  In each panel the solid curve is $M_+(p^2)$; the dashed curve is $M_-(p^2)$; and the dotted curve is $M_W(p^2)$.  All results obtained with $\omega=0.4\,$GeV.}
\end{figure}

\begin{figure}[t]
\centerline{\includegraphics[width=0.53\textwidth]{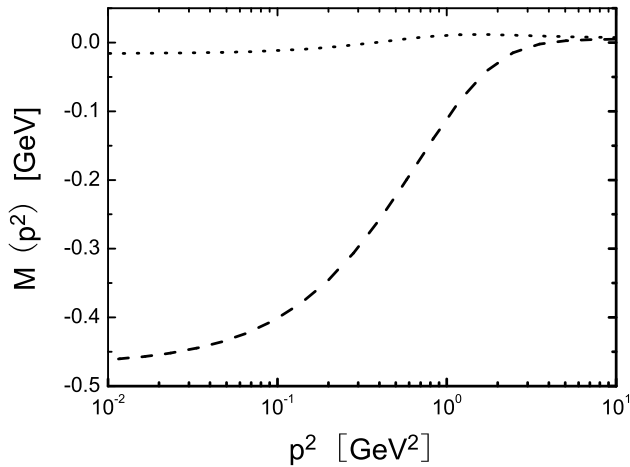}}

\caption{\label{figMpmp2B} Momentum dependence of the dressed-quark mass-function: \textit{dashed curve} -- $M_-(p^2)$; \textit{dotted curve} -- $M_W(p^2)$, both obtained with $m(\zeta_{19})=5\,$MeV and $\omega=0.4\,$GeV.}
\end{figure}

It is noteworthy that we also find a Wigner solution when applying Eq.\,(\ref{renormS}); i.e., in this case, for nonzero current-quark mass we find an analogue of $M_W$ in Eq.\,(\ref{gapsolnjl}).  This suggests that the presence of the Wigner solution is not contingent upon the pointwise behaviour of the gap equation's kernel and supports an interpretation of our findings in the context of QCD.  
(Indeed, all of the features identified herein are also expressed, e.g., in the model of Ref.\,\cite{watson}.)
We observe that in the model of this section the Wigner solution is naturally represented by two momentum-dependent functions; namely, $A_W(p^2,\zeta^2)$, $B_W(p^2,\zeta^2)$.  As indicated by Figs.\,\ref{MTAB}, \ref{figMpmp2}, \ref{figMpmp2B}, at $\hat m = \hat m_{\rm cr}$ these functions meld pointwise with their counterparts in the DCSB $M_-(p^2)$ solution.   

We emphasise that all solutions of the gap equation evolve smoothly with current-quark mass on the domain specified in Eq.\,(\ref{MTmcr}).  Moreover, the Nambu solution characterised by $M_+(p^2)$ evolves smoothly for all values of the current-quark mass.  Naturally, this does not necessarily entail that it's pointwise behaviour at a given value of current-quark mass is obtainable via a perturbative expansion in $\hat m$ about the nonzero DCSB result at $\hat m = 0$, which is essentially nonperturbative in the coupling.  Indeed, the commonality of behaviour between the two models we discuss explicitly herein, and with the others we have considered, suggests strongly that $\hat m_{\rm cr}$ is the radius of convergence for a pointwise chiral expansion of $M_+(p^2)$ around the DCSB chiral-limit result.  However, verification must wait because the method employed in arriving at Eq.\,(\ref{radiusc}) is not workable for the integral gap equation and we do not yet have a tractable alternative.

In considering other rainbow interaction models, we find that $m_{\rm cr}$ assumes similar values in most models that provide a reasonable description of the same low-energy observables.  An exception is the model of Ref.\,\cite{mn83}, in which a solution with $B(s=0)<0$, and $A(s)$ and $B(s)$ continuous on $s\in [0,\infty)$ exists only in the chiral limit, with the interpretation that in this model $m_{\rm cr}=0$.

One might also ask after the effect of dressing the quark-gluon vertex.  As noted above, symmetry ensures that in the chiral limit the gap equation simultaneously admits $M_-(p^2)$ and $M_+(p^2)$ solutions in this instance, too.  The extent of the domain of current-quark mass on which the $M_-(p^2)$ solution persists will probably depend on the structure of the vertex.  We are currently exploring this.

Some may seek an interpretation of the gap equation's distinct solutions.  Here, again, the $\sigma$-model analogy is useful.  Consideration reveals that the gap equation's solutions are representations at an elementary level of the misaligned chiral order parameter that is the focus in discussions of disordered chiral condensates, many consequences of which are reviewed in Ref.\,\cite{dcc}.  We are currently examining others that may be equally useful; e.g., as stated above, by calculating hadron properties in the different vacua.


Equation~(\ref{Mp2uv}) is valid for $M_\pm(p^2)$ and $M_W(p^2)$.  In fact, one can make a stronger statement, while $M_\pm(p^2)$, $M_W(p^2)$ exist:
\begin{equation}
\label{MPMeq}
M_+(p^2)\stackrel{p^2\gg \Lambda_{\rm QCD}^2}{=} M_-(p^2)\stackrel{p^2\gg \Lambda_{\rm QCD}^2}{=}  M_W(p^2)\,.
\end{equation}
This follows from asymptotic freedom, which is a feature of the RGI model and QCD.  One may argue for this result as follows.  On the weak-coupling domain 
\begin{equation}
A_{\pm,W}(p^2,\zeta^2) \approx 1\,,\; \frac{M_{\pm,W}(p^2)}{p^2+M^2_{\pm,W}(p^2)} \approx \frac{M_{\pm,W}(p^2)}{p^2}\,.
\end{equation}
Hence, the gap equation becomes a single linear integral equation for $M_{\pm,W}(p^2)$.  Within the domain on which the preceding steps are valid, that equation can be approximated by a linear second-order ordinary differential equation (d.e.) (e.g., Refs.\,\cite{atkinson,tang,mckellar,munczekde}).  The d.e.\ is the same for every one of the functions $M_{\pm,W}(p^2)$, as is the ultraviolet boundary condition, which is determined by the current-quark mass.  Thus follows Eq.\,(\ref{MPMeq}), a result evident in Figs.\,\ref{figMpmp2}, \ref{figMpmp2B}.


In general the gap equation is solved as a nonlinear integral equation.  It is straightforward to obtain the $M_\pm(p^2)$ solutions via iteration.  However, for nonzero current-quark mass the Wigner solution is harder to fix.  A careful seed function for iteration must be chosen and an adaptive iterative approach employed to reach the solution.  To be concrete, a Chebyshev expansion was employed for the solution functions and a Newton iteration procedure used to determine the coefficients, after the manner of Ref.\,\cite{JCRBthesis}.  The seed for iteration was $A=1$ and $B= - 10 \,m(\zeta)$, and convergence to solution first obtained with $\omega D = (0.6\,{\rm GeV})^3$.  With $\omega=0.4\,$GeV, small steps in $D$ were subsequently made to reach the value in Eq.\,(\ref{omegaD}).  All solutions are illustrated in Fig.\,\ref{figMpmp2} and, for clarity and emphasis, we compare $M_-(p^2)$ and $M_W(p^2)$ in Fig.\,\ref{figMpmp2B}.

In the context of the d.e.\ argument presented above, we add that the model gap equation has often been approximated by a single second-order d.e., which is nonlinear in $M(p^2)$ for infrared momenta but linear in the ultraviolet \cite{atkinson,tang,mckellar,munczekde}.  As we remarked, the ultraviolet boundary condition for all solutions is still fixed by the current-quark mass and the solutions agree.  However, as apparent in Fig.\,\ref{MTAB}, $A_W \not\approx 1$ on the domain of infrared momenta.  Hence, while a single d.e.\ remains a valid approximation for $M_\pm(p^2)$, that is not the case for the Wigner solution.  This emphasises that the differences between $M_+(p^2)$, $M_-(p^2)$ and $M_W(p^2)$ are a primarily infrared effect; i.e., nonperturbative in the coupling.

Using Eq.\,(\ref{MPMeq}) and the subsequent discussion it becomes clear that the gauge-invariant quantity
\begin{equation}
\label{barsigma}
\bar\sigma(m(\zeta)) :=  \lim_{\Lambda\to \infty} 
Z_4(\zeta^2,\Lambda^2)\, N_c \, {\rm tr}_{\rm D}\int^\Lambda_q\! \bar S^{m(\zeta)}(q,\zeta)\,,
\end{equation}
where
\begin{equation}
\bar S^{m(\zeta)}(q,\zeta)  = \frac{1}{2} \left[S_+^{m(\zeta)}(q,\zeta)-S_-^{m(\zeta)}(q,\zeta)\right],
\end{equation}
is a current-quark-mass-dependent quark condensate that is well-defined, finite and unambiguous, and has a perturbative expansion in $\hat m$ on a bounded interval.  In addition, $\bar\sigma(m(\zeta))$ evolves under the renormalisation group in precisely the same manner as the chiral-limit vacuum quark condensate and is identical to the vacuum quark condensate in the chiral limit.  Equation~(\ref{barsigma}) is unique in possessing all these properties.  

\begin{figure}[t]
\centerline{\includegraphics[width=0.54\textwidth]{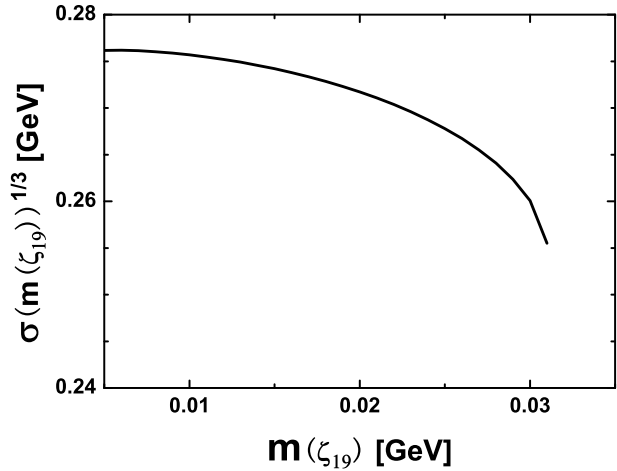}}

\caption{\label{figsm} Evolution with current-quark mass, $m(\zeta_{19})$, of the massive-quark condensate defined in Eq.\,(\protect\ref{barsigma}), calculated with $\omega=0.4\,$GeV and at the renormalisation point $\zeta=19\,$GeV.}
\end{figure}

For additional clarity we remark that for $\hat m\neq 0$ there is no term in a perturbative expansion around the chiral limit of the integrand in Eq.\,(\ref{barsigma}) that is odd in $\hat m$.  Now, since a weak-coupling evaluation of the scalar piece of the quark propagator yields an expression that \emph{is} odd in $\hat m$, then Eq.\,(\ref{barsigma}) contains no term calculable in weak-coupling perturbation theory.  (This may be verified using the d.e.\ analysis discussed above.)  

The behaviour of $\bar\sigma(m(\zeta))$ obtained with the RGI model employed herein is depicted in Fig.\,\ref{figsm}.  As we saw in connection with Fig.\,\ref{figbarM}, here, too, the essentially dynamical component of chiral symmetry breaking decreases with increasing current-quark mass, following the trend predicted by the constituent-quark $\sigma$-term (Ref.\,\cite{HUGS05}, Sec.\,5.2.2).

At this point we can further amplify our interpretation of $\hat m_{\rm cr}$.  A basic ingredient of chiral perturbation theory is the quark condensate because it introduces the mass-scale of DCSB.  If the evolution with current-quark mass of this quantity about its DCSB chiral-limit value cannot be evaluated perturbatively then the expansion in $\hat m$ around $\hat m=0$ has broken down.  The pseudoscalar-meson-loop contribution to this evolution is a long-range piece of the vacuum response.  Now, with Eq.~(\ref{barsigma}), we can evaluate a short-range piece.  While an expansion of both $M_-$ and $M_+$ in $\hat m$ exists, then so does a perturbative expansion of the order parameter $\bar\sigma$.  However, $\bar\sigma$ has no such expansion for $\hat m > \hat m_{\rm cr}$.  NB.\ A chiral expansion is invalid if either the long-range or short-range contribution fails but it is likely that they are linked.

We stress that a straightforward definition of a massive-quark condensate via the trace of a $\hat m \neq 0$ dressed-quark propagator is not useful because it gives a quantity that is quadratically divergent and therefore very difficult to define unambiguously.  The same weakness afflicts the quantity 
\begin{equation}
\check\sigma(m(\zeta),\Lambda) =  Z_4(\zeta^2,\Lambda^2)\, N_c \, {\rm tr}_{\rm D}\int^\Lambda_q\! \check S^{m(\zeta)}(q,\zeta)\,,
\end{equation}
where
\begin{equation}
\label{checkS}
\check S^{m(\zeta)}(q,\zeta)  = \frac{1}{2} \left[S_+^{m(\zeta)}(q,\zeta)+S_-^{m(\zeta)}(q,\zeta)\right].
\end{equation}

The discussion in Refs.\,\cite{mrt98,kurtcondensate,furnstahl} provides a context for Eqs.\,(\ref{barsigma}) -- (\ref{checkS}), and a connection between our reasoning and that used in other frameworks to calculate the quark condensate. 

\section{Summary}
On a bounded interval of current-quark mass, ${\cal D}(\hat m)=\{\hat m \; | \;0 \leq \hat m < \hat m_{\rm cr} \}$, realistic models of QCD's gap equation can simultaneously admit two inequivalent dynamical chiral symmetry breaking (DCSB) solutions for the dressed-quark mass-function, $M_\pm(p^2)$, and a solution that is unambiguously connected with the realisation of chiral symmetry in the Wigner mode, $M_W(p^2)$.  The DCSB solutions are distinguished by their value at the origin: $M_+(p^2=0)>0$ and $M_-(p^2=0)<0$.  In the ultraviolet all three solutions coincide with the running current-quark mass.  

The pointwise values of all solutions evolve continuously with current-quark mass within ${\cal D}(\hat m)$.  
However, things change at the upper boundary.  The $M_W$ solution, whose chiral limit value is perturbative in the coupling, becomes identical to the essentially nonperturbative solution $M_-$ that is actually characteristic of DCSB.  Moreover, both disappear for $\hat m > \hat m_{\rm cr}$, a domain whereupon the current-quark mass is large enough to completely destabilise these solutions.  Only the positive $M_+$ solution exists on this domain.  Furthermore, we provided evidence that the upper boundary of ${\cal D}(\hat m)$ also defines the radius of convergence for an expansion of $M_+$ in current-quark mass around its DCSB chiral-limit form.

Thus one has the coalescing of two qualitatively distinct solutions at $\hat m_{\rm cr}$, the persistence of only one essentially nonperturbative solution for $\hat m > \hat m_{\rm cr}$ plus the breakdown of a chiral expansion for this solution, and the simultaneous loss of a well-defined mass-dependent quark condensate.  
This behaviour supports an interpretation of $\hat m_{\rm cr}$ as the upper bound on the domain within which a perturbative expansion of physical quantities in the current-quark mass around their chiral-limit values can uniformly be valid.  
In a phenomenologically efficacious renormalisation-group-improved rainbow-ladder truncation of QCD's Dyson-Schwinger equations, the critical current-quark mass corresponds to a mass $m_{0^-} \sim 0.45\,$GeV (\mbox{$m_{0^-}^2 \sim 0.20\,$GeV$^2$}) for a pseudoscalar meson constituted of equal mass current-quarks \cite{andreaspi}.  NB.\ Irrespective of the current-mass of the other constituent, a meson containing one current-quark whose mass exceeds $\hat m_{\rm cr}$ is never within the domain of uniform convergence.

A value of similar magnitude was deduced in Refs.\,\cite{awtconverge,young} as the scale below which accuracy may be expected from the approximation of observables through a perturbative expansion in pion-like pseudoscalar-meson mass.  This scale also marks the boundary below which observables should exhibit that curvature as a function of pion-like pseudoscalar-meson mass which is characteristic of chiral effective theories.

\begin{acknowledgments}
Lei Chang and Yu-Xin Liu acknowledge support from Argonne National Laboratory, and are grateful for the hospitality extended by the Theory Group during a visit when part of this work was performed. 
We acknowledge profitable interactions with 
P.~Jaikumar, D.~Nicmorus and P.\,C.\ Tandy.
%
This work was supported by: Department of Energy, Office of Nuclear Physics, contract no.\ DE-AC02-06CH11357;
the National Natural Science Foundation of China under contract nos.\ 10425521, 10075002 and 10135030;
the Major State Basic Research Development Program of China under contract no.\
G2000077400;
and the Doctoral Program Foundation of the Ministry of Education, China, under grant No.\ 20040001010.
One of the authors (Y.-X.~Liu) would also like to acknowledge support from the Foundation for University Key Teacher by the Ministry of Education, China.
\end{acknowledgments}


\begin{thebibliography}{40}

\bibitem{lane} K.\,D.\ Lane,
Phys.\ Rev.\ D {\bf 10}, 2605 (1974).

\bibitem{politzer} H.\,D.\ Politzer,
Nucl.\ Phys.\ B {\bf 117}, 397 (1976). 

\bibitem{cdragw} C.\,D.~Roberts and A.\,G.~Williams,
  Prog.\ Part.\ Nucl.\ Phys.\  {\bf 33}, 477 (1994). 

\bibitem{bowman2} P.\,O.~Bowman, U.\,M.~Heller, D.\,B~Leinweber and A\,G.~Williams, 
  Nucl.\ Phys.\ Proc.\ Suppl.\ \textbf{119}, 323 (2003).

\bibitem{bowman} J.\,B.~Zhang, P.\,O.~Bowman, R.\,J.~Coad, U.\,M.~Heller, D.\,B.~Leinweber and A.\,G.~Williams,
Phys.\ Rev.\ D {\bf 71}, 014501 (2005).

\bibitem{mrt98} P.\,Maris, C.\,D.\ Roberts and P.\,C.\ Tandy,
Phys.\ Lett.\ \textbf{B\,420}, 267 (1998).

\bibitem{pennington} M.\,R.~Pennington,
  Nucl.\ Phys.\ A {\bf 623}, 189C (1997).

\bibitem{weisenjl} U.~Vogl and W.~Weise,
Prog.\ Part.\ Nucl.\ Phys.\  \textbf{27}, 195 (1991). 

\bibitem{klevanskynjl} S.\,P.~Klevansky, 
Rev.\ Mod.\ Phys.\  \textbf{64}, 649 (1992). 
  
\bibitem{ebertnjl}
  D.~Ebert, H.~Reinhardt and M.~K.~Volkov,
  Prog.\ Part.\ Nucl.\ Phys.\  \textbf{33}, 1 (1994). 
  
\bibitem{tandyrev} P.\,C.~Tandy,
  Prog.\ Part.\ Nucl.\ Phys.\  {\bf 39}, 117 (1997).

\bibitem{cahillrev} R.\,T.~Cahill and S.\,M.~Gunner,
  Fizika \textbf{B\,7}, 171 (1998).
  
\bibitem{schmidtrev} C.\,D.\ Roberts and S.\,M.\ Schmidt,
Prog.\ Part.\ Nucl.\ Phys.\  \textbf{45}, S1 (2000). 

\bibitem{alkoferrev} R.~Alkofer and L.~von~Smekal, 
Phys.\ Rept.\ \textbf{353}, 281 (2001). 

\bibitem{marisrev} P.\ Maris and C.\,D.\ Roberts,
Int.\ J.\ Mod.\ Phys.\ \textbf{E\,12}, 297 (2003). 

\bibitem{fnEuclidean} We use a Euclidean metric, with:  $\{\gamma_\mu,\gamma_\nu\} = 2\delta_{\mu\nu}$; $\gamma_\mu^\dagger = \gamma_\mu$; $a \cdot b = \sum_{i=1}^4 a_i b_i$; and ${\rm tr}[\gamma_5\gamma_\mu\gamma_\nu\gamma_\rho\gamma_\sigma]= 
-4\,\epsilon_{\mu\nu\rho\sigma}\,, \epsilon_{1234}= 1\,.$  For a spacelike vector $k_\mu$, $k^2>0$.

\bibitem{mr97} P.~Maris and C.D.~Roberts, Phys.\ Rev.\ C {\bf 56}, 3369 
(1997). 

\bibitem{fnpressure} The pressure is defined as the negative of the effective-action.  Hence, the effective-action difference is zero between two vacuum configurations of equal pressure.  A system's ground state is that configuration for which the pressure is a global maximum or, equivalently, the effective-action is a global minimum.  An elucidation may be found in: R.\,W.~Haymaker,
  Riv.\ Nuovo Cim.\  \textbf{14}, No.\,8, 1 (1991).

\bibitem{reggcm} R.\,T.~Cahill and C.\,D.~Roberts,
  Phys.\ Rev.\ D {\bf 32}, 2419 (1985).
  
\bibitem{cdrqed} C.\,D.~Roberts and R.\,T.~Cahill,
  Phys.\ Rev.\ D {\bf 33}, 1755 (1986).

\bibitem{fn1} This form for the gluon two-point function implements a four-dimensional-cutoff version of the Nambu--Jona-Lasinio model.  Typically \cite{klevanskynjl}, $\tilde\Lambda \sim 1\,$GeV provides a reasonable phenomenology.

\bibitem{HUGS05} A.~H\"oll, C.\,D.~Roberts and S.\,V.~Wright, ``Hadron physics and Dyson-Schwinger equations,'' nucl-th/0601071, to appear in proceedings of \textit{20th Annual Hampton University Graduate Studies Program (HUGS 2005)}, Newport News, VA, 31\,May -- 17\,Jun 2005. 

\bibitem{hatsuda} T.~Hatsuda,
  Phys.\ Rev.\ Lett.\  {\bf 65}, 543 (1990).

\bibitem{munczek} H.\,J.~Munczek,
  Phys.\ Rev.\ \textbf{D\,52} 4736 (1995).
  
\bibitem{truncscheme} A.~Bender, C.\,D.~Roberts and L.~von Smekal, 
Phys.\ Lett.\ {\bf B 380}, 7 (1996). 

\bibitem{detmoldvertex} A.~Bender, W.~Detmold, C.\,D.~Roberts and A.\,W.~Thomas, 
Phys.\ Rev.\ \textbf{C\,65}, 065203 (2002). 

\bibitem{bhagwatvertex} M.\,S.~Bhagwat, A.~H\"oll, A.~Krassnigg, C.\,D.~Roberts and P.\,C.~Tandy,
Phys.\ Rev.\ {\bf C\,70}, 035205 (2004). 

\bibitem{jainmunczek} P.~Jain and H.\,J.~Munczek,
  Phys.\ Rev.\ {\bf D\,48}, 5403 (1993). 

\bibitem{klabucar} D.~Klabu\v{c}ar and D.~Kekez,
  Phys.\ Rev.\ {\bf D\,58}, 096003 (1998). 

\bibitem{maristandy1} P.~Maris and P.\,C.~Tandy,
  Phys.\ Rev.\ {\bf C\,60}, 055214 (1999). 
  
\bibitem{mn83} H.\,J.~Munczek and A.\,M.~Nemirovsky, Phys.\ Rev.\ {\bf D 28} 
(1983) 181. 

\bibitem{mm97} D.\,W.\ McKay and H.\,J.\ Munczek,
Phys.\ Rev.\ D {\bf 55}, 2455 (1997).

\bibitem{richardson} J.\,L.\ Richardson,
Phys.\ Lett.\ B {\bf 82}, 272 (1979).

\bibitem{raya3} P.~Maris, A.~Raya, C.\,D.~Roberts and S.\,M.~Schmidt,
  Eur.\ Phys.\ J.\ {\bf A\,18}, 231 (2003).
  
\bibitem{bhagwat} M.S.~Bhagwat, M.A.~Pichowsky, C.D.~Roberts and P.C.~Tandy,
Phys.\ Rev.\ {\bf C\,68}, 015203 (2003). 

\bibitem{bhagwat2} M.\,S.~Bhagwat and P.\,C.~Tandy,
Phys.\ Rev.\ {\bf D\,70}, 094039 (2004). 

\bibitem{alkoferdetmold} R.~Alkofer, W.~Detmold, C.\,S.~Fischer and P.~Maris,
  Nucl.\ Phys.\ Proc.\ Suppl.\  {\bf 141}, 122 (2005). 

\bibitem{empirad} A.~H\"oll, A.~Krassnigg, P.~Maris, C.\,D.~Roberts and S.\,V.~Wright,
  Phys.\ Rev.\ C {\bf 71}, 065204 (2005).
  
\bibitem{watson} R.~Alkofer, P.~Watson and H.~Weigel,
  Phys.\ Rev.\ D {\bf 65}, 094026 (2002).
  
\bibitem{dcc} B.~Mohanty and J.~Serreau,
  Phys.\ Rept.\  {\bf 414}, 263 (2005).
  
\bibitem{atkinson} D.~Atkinson and P.\,W.~Johnson,
  Phys.\ Rev.\ D {\bf 37}, 2290 (1988);
  \textit{ibid}., 2296 (1988).
  
\bibitem{tang} G.~Krein, P.~Tang and A.\,G.~Williams,
  Phys.\ Lett.\ B {\bf 215}, 145 (1988).

\bibitem{mckellar} C.\,D.~Roberts and B.\,H.\,J.~McKellar,
  Phys.\ Rev.\ D {\bf 41}, 672 (1990).

\bibitem{munczekde} H.\,J.~Munczek and D.\,W.~McKay,
  Phys.\ Rev.\ D {\bf 42}, 3548 (1990).
  
\bibitem{JCRBthesis} J.\,C.\,R.~Bloch, ``Numerical investigation of fermion mass generation in QED,'' hep-ph/0208074.
  
\bibitem{kurtcondensate} K.~Langfeld, H.~Markum, R.~Pullirsch, C.\,D.~Roberts and S.\,M.~Schmidt,
  Phys.\ Rev.\ C {\bf 67}, 065206 (2003).
  
\bibitem{furnstahl} H.-B.~Tang and R.\,J.~Furnstahl, ``The Gluon condensate and running coupling of QCD,'' hep-ph/9502326.

\bibitem{andreaspi} A.~Krassnigg and P.~Maris,
  J.\ Phys.\ Conf.\ Ser.\  \textbf{9}, 153 (2005).

\bibitem{awtconverge} A.\,W.~Thomas,
  Nucl.\ Phys.\ Proc.\ Suppl.\  {\bf 119}, 50 (2003).
  
\bibitem{young} R.\,D.~Young, D.\,B.~Leinweber and A.\,W.~Thomas,
  Prog.\ Part.\ Nucl.\ Phys.\  {\bf 50}, 399 (2003).
  
\end{thebibliography}
\end{document}